\documentclass[12pt,preprint]{aastex}

\keywords{atomic data- atomic processes- line: identification- ISM:
  lines and bands - Galaxy: halo- quasars: absorption lines - X-rays:
  general }

\usepackage[final]{pdfpages}
\usepackage{epstopdf, graphicx, natbib}
\bibpunct{(}{)}{,}{a}{}{;}
\usepackage{times}
\usepackage{rotating,lscape}
\usepackage{epsfig}
\usepackage{graphics,graphicx}
\usepackage{epstopdf}

\def\chandra{{\it Chandra}~}
\def\xmm{{\it XMM-Newton}~}

\def\chandran{{\it Chandra}}
\def\xmmn{{\it XMM-Newton}}

\newcommand{\ignore}[1]{}

\def\lya{\ifmmode {\rm Ly}\alpha~ \else Ly$\alpha$~\fi}
\def\lyan{\ifmmode {\rm Ly}\alpha \else Ly$\alpha$\fi}
\def\lyb{\ifmmode {\rm Ly}\beta~ \else Ly$\beta$~\fi}
\def\lyg{\ifmmode {\rm Ly}\gamma~ \else Ly$\gamma$~\fi}
\def\civ{\ifmmode {\rm C}\,{\sc iv}~ \else C\,{\sc iv}~\fi}
\def\civn{\ifmmode {\rm C}\,{\sc iv}~ \else C\,{\sc iv}\fi}
\def\cvi{\ifmmode {\rm C}\,{\sc vi}~ \else C\,{\sc vi}~\fi}
\def\cvin{\ifmmode {\rm C}\,{\sc vi} \else C\,{\sc vi}\fi}

\def\oii{{{\rm O}\,\hbox{{\sc ii}}~}}
\def\oiin{{{\rm O}\,\hbox{{\sc ii}}}}

\def\oiiin{{{\rm O}\,{\sc iii}}}

\def\oivn{{{\rm O}\,{\sc iv}}}
\def\ov{{{\rm O}\,{\sc v}~}}

\def\ovi{{{\rm O}\,{\sc vi}~}}
\def\ovin{{{\rm O}\,{\sc vi}}}

\def\ovin{{{\rm O}\,{\sc vi}}}
\def\oviin{{{\rm O}\,{\sc vii}}}

\def\gax{${_>\atop^{\sim}}$}
\def\lax{${_<\atop^{\sim}}$}

\def\etal   {{\it et~al.}~}

\def\kalpha{{K$\alpha$~}}
\def\kalphan{{K$\alpha$}}
\def\kbeta{{K$\beta$~}}

\title{The \ovi mystery: mismatch between X-ray and UV column densities}

 \author{
 S.~Mathur\altaffilmark{1,2}, 
 F. Nicastro\altaffilmark{3,4}, 
 A.~Gupta\altaffilmark{5}, 
Y.~Krongold\altaffilmark{6}, 
B.M.  McLaughlin\altaffilmark{7,8}, 
N. Brickhouse\altaffilmark{4}, 
A. Pradhan\altaffilmark{1} 
 } 

 \altaffiltext{1}{\ignore{OSU}Department of Astronomy, The Ohio State
   University, 140 W 18th Ave, Columbus, OH 43210, USA}
 \altaffiltext{2}{\ignore{CCAPP}Center for Cosmology and AstroParticle
   Physics, The Ohio State University, 191 West Woodruff Ave, Columbus,
   OH 43210, USA} \altaffiltext{3}{\ignore{INAF}OsservatorioAstronomico
   di Roma --INAF, Via di Frascati 33, I-00040 Monte Porzio Catone, RM,
   Italy } \altaffiltext{4}{\ignore{CfA}Harvard-Smithsonian Center for
   Astrophysics, 60 Garden Street, Cambridge, MA 02138, USA }
 \altaffiltext{5}{\ignore{ColsState}Department of Biological and
   Physical Sciences, Columbus State Community College, Columbus, OH
   43215, USA} 
 \altaffiltext{6}{\ignore{UNAM}Instituto de Astronomia, Universidad
   Nacional Autonoma de Mexico, Cuidad de Mexico, Mexico}
\altaffiltext{7}{\ignore{CTAMOP}Center for theoretical atomic and
  molecular physics (CTAMOP), School of mathematics and physics, Queen's
  University Belfast, Belfast B17 INN, UK}
\altaffiltext{8}{\ignore{ITAMP}Institute for theoretical atomic and
  molecular physics (ITAMP), Harvard-Smithsonian Center for
  Astrophysics, MS-14, 60 Garden Street, Cambridge, MA 02138, USA}

 \email{smita@astronomy.ohio-state.edu}

\begin{document}
   
\begin{abstract}
  The UV spectra of Galactic and extragalactic sightlines often show
  \ovi absorption lines at a range of redshifts, and from a variety of
  sources from the Galactic circumgalactic medium to AGN outflows. Inner
  shell \ovi absorption is also observed in X-ray spectra (at
  $\lambda=22.03$ \AA), but the column density inferred from the X-ray
  line was consistently larger than that from the UV
  line. Here we present a solution to this discrepancy for the $z=0$
  systems. The \oii \kbeta line $^4S^0 \rightarrow (^3D)3p ^4P$ at
  $562.40$ eV ($\equiv$22.04\AA) is blended with the \ovi \kalpha line
  in X-ray spectra. We estimate the strength of this \oii line in two
  different ways and show that in most cases the \oii line accounts for
  the entire blended line. The small amount of \ovi equivalent width
  present in some cases has column density entirely consistent with the
  UV value. This solution to the \ovi discrepancy, however, does not
  apply to the high column density systems like AGN outflows. We discuss
  other possible causes to explain their UV/X-ray mismatch. The \ovi and
  \oii lines will be resolved by gratings on-board the proposed mission
  {\it Arcus} and the concept mission {\it Lynx} and would allow
  detection of weak \ovi lines not just at $z=0$ but also at higher
  redshift.

\end{abstract}

\section{Introduction}

\ovi absorption at $\lambda\lambda 1032, 1038$ due to the transitions
$1s^22s\rightarrow 1s^22p$ is ubiquitous in all far-ultraviolet (FUV)
spectra of Galactic and extragalactic sightlines observed with the Far
Ultraviolet Space Explorer ($\it FUSE$ with resolution R$=
\lambda/\Delta \lambda$\gax$20,000$; e.g Sembach \etal 2003). Redshifted
\ovi absorption is also observed in UV spectra of extragalactic
sightlines observed with the Hubble Space telescope ($\it HST$; with R
up to $\approx 46,000$, e.g. Crenshaw et al. 1999). The \ovi absorption
lines are observed in several active galactic nuclei (AGN) associated
systems, intervening systems, Galactic high velocity clouds, and from
the thick disk of the Galaxy.  The inner shell (n=1) transition of \ovi
($1s^2 2s \rightarrow 1s2s2p$ at $\lambda 22.03$\AA) lies in the soft
X-ray band, so the \ovi column density can be probed with X-ray
spectroscopy as well. \chandra and \xmm can perform high resolution
grating spectroscopy with R$\approx 500$, and several grating spectra
with good signal-to-noise ratio (S/N) show extragalactic and Galactic
absorption from \ovi (e.g. Williams et al. 2005). Normally, transitions
from Li-like ions of abundant low-Z elements are observed in the UV and
from H- and He-like ions in the X-rays. The Li-like ion \ovi is quite
unique with lines detected both in UV and X-ray bands.

Whatever the origin of \ovi absorbers, one thing should be quite clear:
since both the UV and the X-ray transitions arise from the \ovi $1s^2
2s$ ground state, the column densities derived from both lines {\it
  must} match and should reflect the number of \ovi ions in the ground
state.\footnote{as long as the background continuum source is the
  same. See \S 2, point 4.}  The beauty of absorption line physics is
such that it does not matter if the absorbing gas is multi-phase; we
observe the entire column density of the ion (\ovi in this case) along
the line of sight (at a given redshift). Given the much larger
resolution of UV spectrographs, the UV \ovi absorption lines are often
resolved into multiple velocity components, but their integrated column
density must match the X-ray \ovi column density. Instead, the UV column
densities are repeatedly observed to be smaller by factors of several
(up to $\sim 7$; Arav et al. 2003).

This \ovi discrepancy is an important problem to worry about for several
reasons. First of all, the \ovi discrepancy is observed in a variety of
systems: e.g. (1) AGNs: The associated \ovi X-ray line in the Seyfert
galaxy NGC~5548 was found to have substantially higher column density
than the \ovi UV line (Arav \etal 2003). The \ovi problem is likely to
be common to all AGN outflows, but gets noticed only when the \ovi X-ray
line is strong enough for a detection. (2) The Galaxy interstellar
medium (ISM)/ circumgalactic medium (CGM): The $z=0$ absorption system
toward Mrk 421 shows a clear mismatch between X-ray and UV \ovi column
densities.  Thus the \ovi problem may be generic to all astrophysical
systems of warm/hot plasma, from hot stars, the interstellar medium of
galaxies, and AGN outflows, to the Galactic corona \& intergalactic
medium (IGM). Resolving the \ovi problem is also important for
astrophysical interpretation, e.g. for understanding the physical
conditions in the absorbing plasma. The ratio of the \oviin/\ovi column
density is a powerful diagnostic of gas temperature (e.g. Mathur,
Weinberg \& Chen, 2003), so knowing the correct \ovi column density is
crucial. Understanding inner shell transitions would be crucial in cases
where UV observations are unavailable (e.g. currently at $z=0$ and in
the era beyond $HST$ for higher redshifts), when the X-ray line would be
the only road to \ovin. Resolving the \ovi issue is of great importance
to astronomy in general, X-ray astronomy in particular, and possibly to
atomic physics.

In \S 2 we discuss possible causes of the \ovi discrepancy. In \S 3 we
propose a solution to the $z=0$ systems and conclude in \S 4.

\section {Possible causes of the \ovi discrepancy}  

In this section we discuss several possible causes of the \ovi discrepancy. 

1. Data quality. The quality of the X-ray grating spectra is not as high
as the UV spectra, both in terms of S/N and spectral resolution. While
the \ovi UV absorption lines are well resolved in FUSE/HST spectra, the
\ovi X-ray lines are unresolved by \chandra and \xmmn.  Saturation
effects are also better understood in the UV because of the doublet
nature of the line. As a result, errors on the equivalent width of the
UV absorption lines are much smaller than those on the X-ray lines,
which get transferred to the errors on the \ovi column densities.  Weak
(small column density) lines can be detected in the UV, but not by current
X-ray gratings.  However, the nature of the UV/X-ray discrepancy is
opposite to this expectation based on data quality: X-ray lines are
actually observed to be far {\it stronger} than expected from UV \ovi
measurements. It is thus highly unlikely that the poorer data quality of
X-ray spectra contribute much to the \ovi problem. \xmm detection of the
z=0 \ovi absorption line toward Mrk 421 is robust, with a $3.7\sigma$
sigma detection (EW$=3.3\pm0.9$ m\AA, $1\sigma$ error, Rasmussen et
al. 2007). Moreover, the line was also independently detected by
\chandra with similar strength (EW$=2.4\pm0.9$ m\AA, Williams et
al. 2005), a $2.7\sigma$ detection, increasing the combined significance
to $4.5\sigma$. Thus, we cannot just dismiss this robust detection as a
statistical accident.

2. Saturation. If UV \ovi lines are saturated, the column density might
be underestimated (Arav et al. 2003).  For the unresolved lines,
  the observed EW is related both to the column density and the velocity
  dispersion parameter (or the Doppler parameter {\it b}). The column
  density and the {\it b}-parameter can be disentangled by using
  different transitions of the same ion, as we did in Williams et
  al. (2005) for the $z=0$ absorption along the Mrk 421 sightline. Using
  the UV \ovi absorption line doublet, we constrained the {\it
    b-}parameter tightly and found that for this value of {\it b}, the
  X-ray \ovi line has column density about 0.5 dex higher than the UV
  line.  If the X-ray \ovi line is saturated, the column density would
  be even higher.  Thus the \ovi discrepancy cannot be attributed to
simple phenomena such as saturation in the $z=0$ systems, so the
solution has to lie elsewhere, and may be different for different
systems.

3. Variability. Another possibility is variability, discussed by Arav et
al. (2003). AGN absorption lines are known to vary, so if UV and X-ray
measurements are not obtained at the same time, the measured column
densities may be different. However, variability cannot be the cause of
the \ovi discrepancy in CGM/IGM systems which do not vary.

4. Size of the emission region. In AGNs, the size of the X-ray emission
region is known to be smaller than the UV continuum size. Therefore, if
part of the \ovi X-ray absorption happens inside the UV continuum
region, and both X-ray and UV \ovi absorption takes place outside the UV
emitting region, there might be a discrepancy, with X-ray \ovi
absorption probing a larger column density. Once again this
cannot be the cause of the \ovi discrepancy in CGM/ISM systems which are
far from the background quasars.

5. Excitation. In a high temperature plasma, $2s\rightarrow 2p$ electron
impact excitation and recombination may suppress absorption from the
ground level and contribute to emission. However, the Einstein A$_{21}$
coefficient for the 2p--2s decay is extremely large ($4.2\times 10^8
s^{-1}$), which means any excited \ovi atom decays immediately to the
ground state.  Even in a plasma with an unrealistically high
collision/photoexcitation rate ($>>10^8/sec$), the standard Boltzmann
equation only allows for the ground state to be depleted by a factor of
one-half, and the observed discrepancy is much larger than this.

6. Atomic physics. The wavelengths, oscillator strengths,
photoionization cross sections of the UV \ovi $\lambda\lambda 1032,
1038$ doublet are well known and have been established for
years. However, the same cannot be said about the \ovi X-ray line.
Pradhan (2000) showed that the wavelength of the inner shell transition
($1s^2 2s \rightarrow 1s2s2p$, also known as the KLL transition because
one K shell electron goes to the L shell) is at 22.05\AA\ and that the
photo-absorption cross section via auto-ionizing resonances 
may be appreciable (with average calculated $f=0.576$). It is possible
that absorption strengths are not accurate.  However, \chandra
observations have shown that the wavelength of the KLL absorption line
matches exactly the calculated value (Kaastra et al. 2000). Laboratory
experiments have also confirmed the line wavelength, though in emission
(22.02$\pm 0.002$, Schmidt \etal 2004). Nevertheless, calculations of
photoabsorption resonance oscillator strengths are far more complex than
wavelength calculations.  Resulting from a resonance transition, the
\ovi KLL absorption cross section is already extremely large.  If the
\ovi discrepancy is due to a miscalculation of the X-ray absorption
strength, the correct value would have to be several times larger than
any other known inner-shell transition in this wavelength
region. Subsequent calculations by Behar \& Kahn (2002) resulted in
$f=0.525$, not too different from that of Pradhan (2000).  Thus there is
no obvious atomic physics explanation to resolve the \ovi problem.

\subsection {\oii contamination}

A possible solution has emerged from theoretical and experimental work
on inner shell transitions of oxygen ions (\oii and \oiiin, Bizau et
al. 2015; \oivn, McLaughlin et al. 2015; \ov and \ovin, McLaughlin et
al. 2017). The most recent data on \ovi are provided by McLaughlin et
al. (2017) who provide theoretical as well as experimental values for
the inner shell transition of the X-ray \ovi line ($1s^2 2s \rightarrow
1s2s2p$) wavelength (22.032 \AA) and oscillator strength ($f=0.328\pm
0.05$ from experiments and $f=0.387$ from theory; their Table
5). Including the effect of radiation damping, the effective oscillator
strength is $f=0.49$.  The inner shell \oii transitions are discussed in
Bizau et al. (2015). Experimental and theoretical values of line
wavelengths and oscillator strengths of \oii $1s \rightarrow np$
transitions are given in their Table III. Their ``line-12'', which is a
\kbeta transition $^4S^0 \rightarrow (^3D)3p ^4P$ at $562.40$ eV
($\equiv$22.04\AA) is of interest here ($f=0.038$ from experiment
  and $f=0.022$ from theory. The experimental errors are about
  $15$--$20$\%).

As shown in fig. 5 of McLaughlin et al. (2017), the inner shell \ovi
transition at 22.032 \AA\ (their value) lies very close to an inner
shell \oii transition (line-12) at 22.04 \AA. This difference of about
0.01 \AA\ is below the grating resolutions on board \chandra ($\Delta
\lambda = 0.023$ \AA) and \xmm ($\Delta \lambda = 0.033$ \AA), so the
two lines would be unresolved. The \oii line may therefore contaminate
the \ovi signal. In the following we estimate the degree of this
contamination to figure out whether this would solve the mystery of
excess \ovi column density in X-ray spectra, at least partially; we do
this in two different ways.

\subsubsection {\oii \kbeta ``line 12'' strength from \kalpha}

Let us study the case of Mrk 421 which has one of the highest S/N
\chandra spectra of an extragalactic source. From Williams et
al. (2005) we know that the $z=0$, \ovi X-ray line is detected in this
source with EW$=2.4\pm 0.9$ m\AA. The corresponding column density is
$\log N(\rm O VI)/ cm^{-2}=15.05^{+0.17}_{-0.22}$. The UV \ovi column
density on the other hand is $\log N(\rm O VI)/cm^{-2}=14.45\pm 0.02$
(including the \ovi high velocity cloud (HVC) column density). Thus
there is a factor of four discrepancy between the X-ray and UV column
densities. The \oii line, however, might make the observed X-ray line EW
artificially high. In the Mrk 421 sightline, we have detected the \oii
\kalpha line with EW$=9.9\pm0.6$ m\AA\ (Nicastro et al. 2016). From this
we estimate the EW of \oii line-12 given the wavelengths of the two
lines and their oscillator strengths (EW$_1$/EW$_2 = (f_1/f_2)
(\lambda_1/\lambda_2)^2 $). Thus the estimated \oii line-12 EW is $1.8$
m\AA, strongly contaminating the \ovi line. The actual \ovi EW is
therefore $2.4-1.8=0.6$ m\AA\ corresponding to column density of
$2.4\times 10^{14}$ cm$^{-2}$ or $\log N(\rm O VI)/cm^{-2}=14.4\pm 0.1$,
entirely consistent with the UV value.

We similarly estimated the degree of \oii contamination in other $z=0$
systems noted in Table 1. It is clear that \oii contamination entirely
accounts for the excess X-ray \ovi column density. In systems like the
$z=0$ absorption toward Mrk 509, the entire line could be the \oii \kbeta
line-12.

\subsubsection {\oii \kbeta ``line 12'' strength from ``line 7''}

Another, and perhaps a better way to estimate the strength of the \oii
\kbeta line-12 contamination to the \ovi line is by comparing the
strengths of the two \oii \kbeta lines: line-12 and line-7. The \oii
\kbeta line-7 (in Bizau et al. 2015) is transition $^4S^0 \rightarrow
(^5S)3p ^4P$ at $555.93$ eV (22.30 \AA).  Given that both line-7 and
line-12 are \kbeta transitions, with similar oscillator strengths (0.03
and 0.04 respectively; from Table III of Bizau et al. 2015), their line
strengths would be similar. In Table 2 we list the EWs of \oii lines in
Galactic sightlines from Nicastro et al. (2016); errors are
  $1\sigma$ and upper limits are $3\sigma$. Because of a bad column in
the \xmm RGS spectrograph, \oii \kalpha line strengths could not be
measured in several spectra (column 2 in Table 2). In column 3 we list
the EW of \oii line-7 and in column 4 we list the EW of the blended line
(\oii line-12 $+$ \ovi \kalphan).  In figure 1 we have plotted these
data for the Galactic sightlines: EW(\oii line-7) vs the blended EW(\oii
line-12)$+$EW(\ovin). Black points are for sightlines where EWs of both
the lines were measured while red points denote one upper-limit and blue
are when both are upper limits (dashed lines). Once again we note that
(EW$_1$/EW$_2 = (f_1/f_2) (\lambda_1/\lambda_2)^2 $). Therefore the EW
ratio of line-12 to line-7 is $1.4\pm 0.6$ (given the oscillator
strengths in Bizau et al. 2015). The blue solid lines in Figure 1
bracket this ratio [0.8; 2.0]. We see that all the points in this plot
lie within the shaded region between the two solid blue lines. This
shows that the contribution of \ovi to the blend (\oii $+$ \ovin) is
minimal and that most of the signal we see is from the \oii line-12, not
from \ovin.

Similar data for extragalactic sightlines are presented in Table 3 and
plotted in Figure 2. Again we see that most of the signal in the blended
line comes from \oii line-12, not from \ovin.  Observations of the
  $z=0$ UV \ovi absorption lines are presented by several authors
  (e.g. Savage et al. 2000; Wakker et al. 2003; Indebetouw \& Shull
  2004; Oegerle et al. 2005; Collins, Shull \& Giroux 2005; Ganguly et
  al. 2005; Savage \& Lehner 2006; Bowen et al. 2008; Welsh \& Lallement
  2008; Barstow et al. 2010; Lehner et al. 2011; Howk \& Consiglio
  2012). These probe the Galactic thin disk, thick disk, halo and the
  high velocity clouds, with UV \ovi column densities at $z=0$ ranging
  from $\approx 3\times 10^{12}$ to \lax $10^{15}$ cm$^{-2}$. The
  corresponding maximum X-ray \ovi EW is $2.09$mA, making insignificant
  contribution to the EWs of the blended lines (\oii line-12 $+$ \ovi
  \kalphan) listed in Tables 2 \& 3.   This shows that what was thought
to be the $z=0$ \ovi \kalpha line is actually \oii \kbeta line-12. The
small UV \ovi column density observed in $z=0$ systems is too small to
make a detectable X-ray line, and the observed data are consistent with
this expectation.

\section{Conclusion}

While we have resolved the \ovi problem at $z=0$ with the \oii blend,
this cannot be the entire solution for intrinsic AGN absorbers at higher
redshift where the observed wavelength of the \ovi X-ray line moves away
from the $z=0$ \oii line. There could be some contamination from \oii in
the host galaxy of the AGN, but this would be negligible given the large
column densities of intrinsic absorbers. Similarly, there could be \oii
contamination from the intrinsic absorber itself, but the high
ionization level of \ovi absorber contains insignificant amount
\oiin. We need to investigate any possible contamination to the
intrinsic absorption on a case by case basis, which we did for the
well-studied AGN NGC~5548. The wavelength of the redshifted X-ray \ovi
line is $22.389$ \AA, which is close to the $z=0$, \ov K$\alpha$ line at
$22.368$ \AA. This $\Delta \lambda = 0.021$ \AA\ is within the \chandra
grating resolution, so the lines are blended. The X-ray \ovi column
density in NGC 5548 is $3.2\pm0.8 \times 10^{16}$ cm$^{-2}$ (Arav et
al. 2003; see also Andrade-Velazquez 2010) and that of UV \ovi is
$4.9\pm 0.6 \times 10^{15}$ cm$^{-2}$. Therefore, the $z=0$, \ov column
density will have to be as large as $8.4\times 10^{15}$ cm$^{-2}$ (given
the \ov oscillator strength of 0.64 from McLaughlin et al. 2017) for
X-ray and UV measurements to match. We do not know the Galactic \ov
column density in the sightline to NGC 5548, but Nevalainen et
al. (2017) have reported Galactic \ov absorption line detection toward
PKS~$2155-304$. The reported EW is $3.0\pm 1.5$ m\AA\ (RGS1) and $3.7\pm
2.3$ m\AA\ (LETG/HRC-S). This corresponds to the column density of
$1\times 10^{15}$ cm$^{-2}$, significantly smaller than what is
required. The intrinsic AGN absorbers thus appear to be complex; while
their X-ray \ovi absorption lines must be blended with other $z=0$ or
host galaxy lines, the degree of contamination is likely insignificant.
For these systems, one or more of the other possibilities discussed in
\S 2, such as saturation, variability, and source size may be
responsible.

Given the strong contamination of the \ovi X-ray line by \oii at z=0,
X-ray measurements of this line require higher spectral resolution, such
as proposed for the {\it Arcus} mission (Smith et al. 2017). With
R$>2500$, the instrumental width at the $22$ \AA\ is $0.009$ \AA,
clearly enough to separate the two lines. Furthermore, with an effective
area of more than $300$ cm$^2$, compared with less than $10$ cm$^2$ for
{\it Chandra}, {\it Arcus} will vastly improve the chances of detecting
the weaker \ovi, not just at $z=0$ but also at higher redshift. A
proposed grating for the {\it Lynx} concept mission (Gaskin et al. 2015)
promises even more capability, with R$>5000$ and an effective area of
more than $4000$ cm$^2$.

\noindent {\bf Acknowledgment:} Support for this work was initially
provided by the National Aeronautics and Space Administration through
Chandra Award Number TM9-0010X to SM issued by the Chandra X-ray
Observatory Center, which is operated by the Smithsonian Astrophysical
Observatory for and on behalf of the National Aeronautics Space
Administration under contract NAS8-03060. SM gratefully acknowledges
support through the Chandra Award Number GO4-1511X as well. FN
acknowledges support through NASA grant number NNX17AD76G. YK acknowledges
support from PAPIIT grant IN-104215. BMMcL acknowledges support by the
US National Science Foundation through a grant to ITAMP at the Harvard
Smithsonian Center for Astrophysics under the visitor's program and
Queen's University Belfast for a visiting research fellowship (VRF).

\clearpage

\begin{table}
\scriptsize
\caption{The $z=0$ systems with X-ray \ovi absorption }
\begin{tabular}{lccccc}
\hline
Sightline & EW (m\AA)   & $\log$ Column density & EW (m\AA)  & $\log$ Corrected column density & $\log$ Column density \\
& (\ovi X-ray) &(\ovi X-ray)  & (\oii line 12)$^c$ &(\ovi X-ray)  &(\ovi UV) \\

\hline
 Mrk $421$ & $2.4\pm0.9^a$ & $15.05^{+0.17}_{-0.22}$ & $1.8$ & $14.4\pm 0.1$ & $14.43\pm 0.02$\\
Mrk $509$ & $4.1\pm 1.4^b$ & $15.4\pm 0.1$ & $5.9\pm1.3$ & $< 14.84^d$ & $14.74^{+0.03}_{-0.05}$\\
PKS $2155-304$ & $< 5.7^e$ & $<15.36$ & $1.5$ & $< 16.0$ & $14.48\pm 0.33$ \\
\hline
\end{tabular}

\noindent
a. Williams et al. 2005. Errors are $1\sigma$. \\
b. Calculated from the column density given in Pinto et al. 2012. \\
c. Calculated from \oii \kalpha EW in Nicastro et al. 2016. Errors are $1\sigma$. \\
d. $1\sigma$ upper limit. \\
e. Williams et al. 2007; $2\sigma$ upper limit.  \\
\end{table}

\begin{table}
\scriptsize
\caption{$z=0$ absorption: Galactic sightlines$^a$ }
\begin{tabular}{lccc}
\hline
Sightline & \oii \kalpha   & \oii \kbeta line-7  & \oii \kbeta (line-12)$+$\ovi \kalpha \\
& EW (m\AA)  & EW (m\AA)  &  EW (m\AA)  \\
\hline
HER X-1 & NA & $< 7$ & $< 12$ \\
PSRB 0833-45 & NA & $8\pm 5$ & $<12$  \\
SAX J1808-3658 & NA & $9\pm3$ & $10\pm 3$  \\
Swift J1753.5-0127 & NA & $< 10$ & $7\pm 3$ \\
EXO 0748-676 & NA & $<9$ & NA  \\
Cyg X-2  & $66\pm 2$ & $<9$ & $8\pm4$  \\
MAXI J0556-332 & $20\pm 5$ & $<12$ & $6\pm 3$ \\
Cyg X-1 & $40\pm2$ & $<15$ & $<16$ \\
Swift J1910.2-0546 & $41\pm3$ & $7\pm 4$ & $15\pm 4$ \\
4U 1636-53 & $87\pm6$ & $9\pm6$ & $20\pm6$  \\
4U 1728-16 &  $40\pm4$ & $22\pm6$ & $8\pm6$  \\
V*V 821 Ara &  $56\pm 2$ & $19\pm 6$ & $25\pm 8$  \\
GS 1826-238 &  NA & $< 25$ & $20\pm9$  \\
HETE J1900.1-2455 & NA & $<23$ & $<22$  \\
4U 2129+12 & NA & $<28$ & $13\pm 9$  \\
4U 1543-624 &  $78\pm 5$ & $20\pm 9$ & $26\pm 9$  \\
Aql X-1 & NA & $29\pm 11$ & $16\pm 9$  \\
4U 1735-444 & NA & $<35 $ & $<15 $  \\
X-Persei & NA & $<43$ & $<39$  \\
XTE J1650-500 & NA & $<61$ & $37\pm 16$  \\
\hline
\hline
\end{tabular}

\noindent
a. The data in Tables 2 and 3 are from \xmm spectra.

\end{table}

\begin{table}
\scriptsize
\caption{$z=0$ absorption: Extragalactic sightlines }
\begin{tabular}{lccc}
\hline
Sightline & \oii \kalpha   & \oii \kbeta line-7  & \oii \kbeta (line-12)$+$\ovi \kalpha  \\
& EW (m\AA)  & EW (m\AA)  &  EW (m\AA)   \\
\hline
1ES 1028+111 & NA & $<25$ & $<28$  \\
1H 0419-577 & NA & $<27$ & $<32$  \\
1H 0707-495 & NA & $6\pm6$ & $10\pm6$ \\
3C 120 & NA & $<26$ & $14\pm 9$ \\
3C 273 & NA & $2\pm 1$ & $5\pm 2 $ \\
3C 390.3 & NA & $23 \pm 15$ & $21\pm 10$ \\
AKN 564 & $42\pm 3$ & $<5$ & $<5$ \\
Ark 120 & NA & $<8$ & $5\pm 4$  \\
E 181+643 & NA & $<80$ & $49\pm 21$  \\
ESO 141-G055 & NA & $<18$ & $<16$  \\
ESO 198 -G24 & NA & $33\pm 18$ & $<51$  \\
ESO 511-G030 & NA & $<41$ & $<33$  \\
Fairall 9 & NA & $17\pm 9$ & $<25$  \\
H 1426+428 & NA & $9\pm 5$ & $7\pm 5$  \\
H 2356-309 & NA & $<19$ & $<19$  \\
HE 1029-1401 & NA & $<68$ & $<69$  \\
IRAS 13224-3809 & NA & $<56$ & $<52$  \\
IRAS 13349+2438 & $<51$ & $<58$ & $28\pm 18$  \\
Mkn 205 & NA & $15\pm 12$ & $23\pm 12$ \\
Mkn 279 & NA & $24\pm 8$ & $<24$  \\
Mkn 421 & NA$^a$ & $3.3\pm 0.5$ & $4.4\pm 0.5$ \\
Mkn 501 & NA & $11\pm 7$ & $9\pm 6$  \\
Mkn 766 & $27\pm 4$ & NA & NA  \\
Mkn 841 & NA & $<36$ & $18\pm 15$  \\
MR 2251-178 & NA & $14\pm 6$ & $12\pm 6$  \\
Mkn 1044 & NA & $<28$ & $<30$ \\
Mkn 335 & NA & $<12$ & $<13$  \\
NGC 4593 & NA & $<36$ & $<28$  \\
NGC 5548 & NA & $38\pm 8$ & $39\pm 9$  \\
NGC 7213 & NA & $<45$ & $<47$ \\
NGC 7469 & NA & $<13$ & $<16$ \\
PG 0804+761 & NA & $<57$ & $<59$ \\
PG 1116+215 & NA & $14\pm 12$ & $23\pm 13$  \\
PG 1211+143 & NA & $37\pm 14$ & $30\pm 15$  \\
PG 1244+026 & NA & $<32$ & $<43$ \\
PG 1553+113 & NA & $<34$ & $<39$ \\
PKS 0548-32 & $23\pm 9$ & $<29$ & $<28$ \\
PKS 0558-504 & NA & $<12$ & $<13$  \\
PKS 2005-489 & NA & $<44$ & $19\pm 16$  \\
PKS 2155-304 & NA$^a$ & $<3$ & $<4$ \\
RE 1034+396 & NA & $<21$ & $<27$ \\
UGC 3973 & NA & $<47$ & $20\pm 15$ \\
Mkn 478 & NA & $<64$ & $41\pm 23$  \\
Mkn 704 & NA & $<60$ & $29\pm 20$  \\
1H 0414+009 & NA & $39\pm 28$ & $<97$ \\
3C 279 & NA & $<90$ & $24\pm 19$ \\
I Zw1 & NA & $<89$ & $<88$ \\
Q 0056-363 & NA & $<79$ & $<78$ \\
S 50716+714 & NA & $32\pm 23$ & $<80$ \\
\hline
\hline
\end{tabular}

\noindent
a. While the line is detected with \chandran, as reported in Table 1, it could
not be measured with \xmmn. \\
\end{table}

\clearpage

\begin{figure}
\begin{center}
\includegraphics[width=10cm]{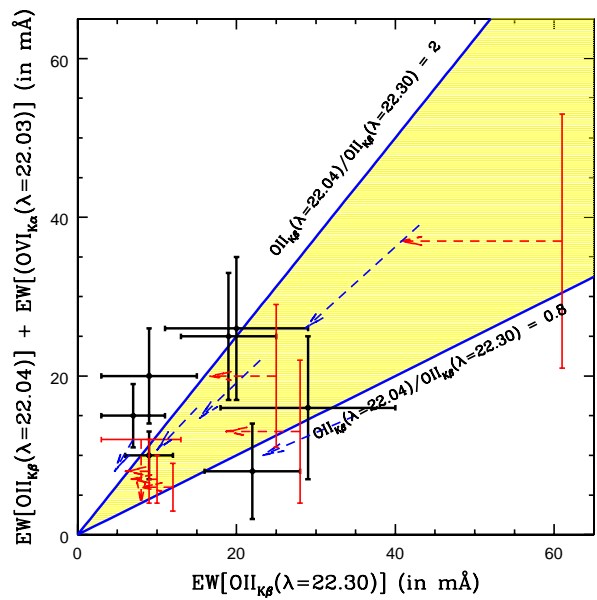}
\end{center}
\caption{The EW of line-7 is plotted vs EW of the blend (line-12$+$\ovi
  \kalpha) at $z=0$ in the Galactic sightlines. Black points are for
  systems where both line-7 and line-12-blend are detected. Red are for
  systems where one of the two lines has only an upper limit (dashed),
  and blue dashed lines are for both the upper limits. The blue solid
  lines bracket the theoretical ratio of EWs of line-7 and line-12. We
  see that all the points are consistent with being in the yellow shaded
  region between the blue lines. This shows that most of the signal in
  the blended line is from line-12, not from \ovi \kalpha.}
\end{figure}

\clearpage

\begin{figure}
\begin{center}
\includegraphics[width=10cm]{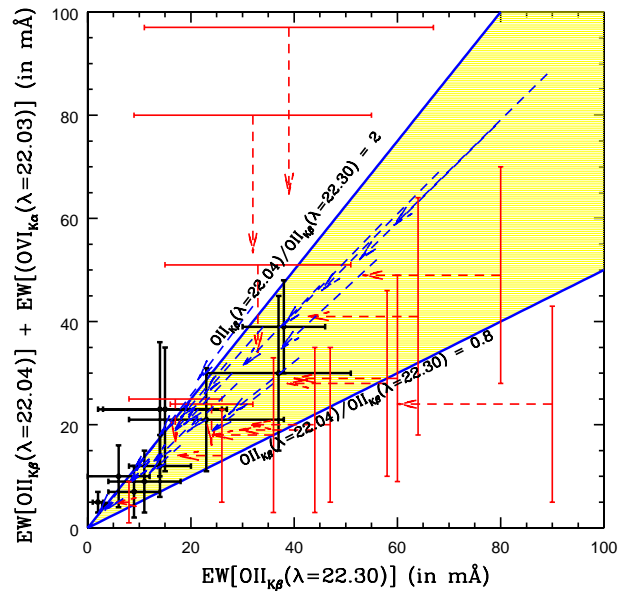}
\end{center}
\caption{Same as in Figure 1, but for extragalactic sightlines. }
\end{figure}


\clearpage

\begin{thebibliography}{}

\bibitem[Andrade-Velazquez et al. 2010]{}Andrade-Velazquez, M.;
  Krongold, Y.; Elvis, M.; Nicastro, F.; Brickhouse, N.; Binette, L.;
  mathur, S.; \& Jimenex-Bailon; E., 2010, ApJ, 711, 888.

\bibitem[Arav et al. 2003]{}Arav, N., Kaastra, J., Steenbrugge, K. et
  al. 2003, ApJ, 590, 174

\bibitem[Barstow et al. 2010]{}Barstow, M., Boyce, D., Welsh, B., et
  al. 2010, ApJ, 723, 1762

\bibitem[Behar \& Kahn 2002]{}Behar, E., and Kahn, S. 2002, in
  Proc. NASA Laboratory Astropysics Workshop, ed. F. Salama (Moffet
  Field: NASA-Ames), 8B

\bibitem[Bizau et al. 2015]{}Bizau, J.-M., Cubaynes, D., Guilbaus, S. et
  al. 2015, Phys. Rev. A., 92, 023401

\bibitem[Bowen et al. 2008]{}Bowen, D., Jenkins, E., Tripp, T., et
  al. 2008, ApJS, 176, 59.

\bibitem[Collins 2005]{}Collins, J., Shull, M., \& Giroux, M., 2005,
  ApJ, 623, 196

\bibitem[Crenshaw et al. 1999]{}Crenshaw, D. Michael; Kraemer, Steven
  B.; Boggess, Albert; Maran, Stephen P.; Mushotzky, Richard F.; \& Wu,
  Chi-Chao, 1999, ApJ, 516, 750

\bibitem[Gaskin et al. 2015]{}Gaskin, J. K., Weisskopf, M. C.,
  Vikhlinin, A., et al., Proc. SPIE, 2015, p. 9601E

\bibitem[Ganguly et al. 2005]{}Ganguly, R., Sembach, K., Tripp, T., \&
  Savage, B., 2005, ApJS, 157, 251

\bibitem[Howk 2012]{}Howk, J. C., \& Consiglio, S., M. 2012, ApJ, 759, 97. 

\bibitem[Indebtouw \& Shull 2004]{}Indebtouw, R. \& Shull, M. 2004, ApJ,
  607, 309

\bibitem[Kaastra et al. 2002]{}Kaastra, J., Steenbrugge, K., Raasen,
  A.J., et al., 2002 A\&A, 386, 427

\bibitem[Lehner et al. 2011]{}Lehner, N., Zech, W., Howk, J., \& Savage,
  B. 2011, ApJ, 727, 46

\bibitem[Mathur et al. 2003]{}Mathur, S., Weinberg, D.H., \& Chen, X.,
  2003, ApJ, 582, 82

\bibitem[McLaughlin et al. 2014]{}McLaughlin, B.-M., Bizau, J.-M.,
  Cubaynes, D., et al. 2014, J. Phys. B: A. Mol. Opt. Phys., 47, 065201

\bibitem[McLaughlin et al. 2017]{}McLaughlin, B.-M., Bizau, J.-M.,
  Cubaynes, D., et al. 2017, MNRAS, 465, 4690

\bibitem[Nevalainen et al. 2017]{}Nevalainen, J.; Wakker, B.; Kaastra,
  J.; Bonamente, M.; Snowden, S.; Paerels, F.; \& de Vries, C., 2017,
  A\&A in press (arXiv:1705.08497)

\bibitem[Nicastro et al. 2016]{}Nicastro, F., Senatore, F., Gupta, A.,
  et al. 2016, MNRAS, 457, 676

\bibitem[Oegerle et al. 2005]{}Oegerle, W., Jenkins, E., Shelton, R.,
  Bowen, D, \& Chayer, P., 2005, ApJ, 622, 377

\bibitem[Pinto et al. 2012]{}Pinto, C., Kriss, G. A., Kaastra, J. S., et
  al. 2012, A\&A, 541, A147

\bibitem[Pradhan 2000]{}Pradhan, A. 2000, ApJL, 454, 165

\bibitem[Rasmussen et al. 2007]{}Rasmussen, A., Kahn, S.M., Paerels, F.,
  et al. 2007, ApJ, 656, 129

\bibitem[Savage et al. 2000]{}Savage, B.D., Sembach, K.R., Jenkins,
  E.D. et al. 2000, ApJ, 538, L27

\bibitem[Savage \& Lehner 2006]{}Savage, B. \& Lehner, N., 2006, ApJS,
  162, 134

\bibitem[Schmidt et al. 2004]{}Schmidt, M. Beiersdorfer, P., Chen, H.,
  et al. 2004, ApJ, 604, 562

\bibitem[Sembach et al. 2003]{}Sembach, K., Wakker, B. P., Savage,
  B. D., et al. 2003, ApJS, 146, 165

\bibitem[Smith et al. 2016]{}Smith, R. K., Abraham, M. H.; Allured,
  R. et al. 2016, Proc. of SPIE, p. 99054M.

\bibitem[Wakker et al. 2003]{}Wakker, B.D., Savage, B.D., Sembach, K.R.,
  et al. 2003, ApJS, 146, 1

\bibitem[Welsh \& Lallement 2008]{}Welsh, B., \& Lallement, R., 2008,
  A\&A, 490, 707

\bibitem[Williams et al. 2005]{}Williams, R.-J., Mathur, S., Nicastro,
  F., et al. 2005, ApJ, 631, 856

\bibitem[Williams et al. 2007]{}Williams, R.-J., Mathur, S., Nicastro,
  F. \& Elvis, M. 2007, ApJ, 665, 247


\bibitem[]{}

\end{thebibliography}
\end{document}